\documentclass[twocolumn,preprintnumbers,amsmath,amssymb,aps,prb]{revtex4-1}
\usepackage{dcolumn}
\usepackage{bm}
\usepackage{subfigure}
\usepackage[version=3]{mhchem}
\usepackage[pdftex]{graphicx}
\usepackage[usenames]{color}
\definecolor{red}{rgb}{1,0,0}
\definecolor{green}{rgb}{0,1,0}
\definecolor{blue}{rgb}{0,0,1}
\definecolor{orange}{rgb}{1,0.5,0}
\definecolor{light-gray}{gray}{0.95}

\include{hyphenation_list}

\graphicspath{{images/}}
\newcommand{\degree}{\ensuremath{^\circ}} 

\usepackage{xspace}
\newcommand{\eg} {e.\,g.\xspace}
\newcommand{\ie} {i.\,e.\xspace}
\newcommand{\etal} {et\,al.\xspace}

\begin{document}
\title{Effects of molecular adsorption on optical losses of Ag~(111) surface. }
\author{A. V. Gavrilenko}
 \email{a.v.gavrilenko@nsu.edu}
\affiliation{Center for Materials Research, Norfolk State University, 700 Park Ave, Norfolk, VA 23504}
\author{C. S. McKinney}
\affiliation{Center for Materials Research, Norfolk State University, 700 Park Ave, Norfolk, VA 23504}
\author{V. I. Gavrilenko}
\affiliation{Center for Materials Research, Norfolk State University, 700 Park Ave, Norfolk, VA 23504}


\begin{abstract}
The first principles density functional theory (DFT) is applied to study effects of molecular adsorption on optical losses of silver (111) surface. The ground states of the systems including water, methanol, and ethanol molecules adsorbed on Ag~(111) surface were obtained by the total energy minimization method within the local density approximation (LDA). Optical functions were calculated within the Random Phase Approximation (RPA) approach. Contribution of the surface states to optical losses was studied by calculations of the dielectric function of bare Ag~(111) surface. Substantial modifications of the real and imaginary parts of the dielectric functions spectra in the near infrared and visible spectral regions, caused by surface states and molecular adsorption, were obtained. The results are discussed in comparison with available experimental data.
\end{abstract}



 \maketitle 

\section{Introduction}\label{sec:intro}  
The effects of metallic surface states and molecular adsorption on the atomic surface structure and electronic properties of transition metals is extensively discussed in literature (see \eg Refs.~\onlinecite{methfessel92,morgenstern02,michaelides06,forster08} and references therein). The silver based metallic nanostructures are actively studied for numerous applications of modern nanotechnology such as nanophotonics and nanoplasmonics \cite{pinchuk04}. It has been demonstrated before that electronic surface states on metallic surfaces substantially contribute to the bonding of physisorbed atoms and molecules \cite{forster08} and to their optical absorption spectra \cite{pinchuk04,persson93}. The semi-classical approach developed by Persson \cite{persson93} using Drude model and modified by Pinchuk \etal \cite{pinchuk04} clearly demonstrated increasingly important contributions of the surfaces states on optical absorption spectra of metallic nanoparticles with descrease of their size. 

Reduction of optical losses in nanostructured materials is one of the most important issues affecting successful engineering of different optoelectronics photonics devices. Optical losses of materials are determined through the imaginary part of the dielectric function ($Im(\varepsilon)$) \cite{saleh07}. Though losses could be compensated through external gain, another possibility is the modification of electronic structure of metallic nanoparticles by chemical adsorption on the surfaces. However, this approach requires detailed understanding of electronic processes accompanied by chemical adsorption on a microscopic scale. 

Recently using the first principle modeling we predicted the effect of the surface states on the dielectric function spectra ($\varepsilon(\omega)$) of the nanometer-thick silver films \cite{zhu08}. It has been shown that a decrease of the silver-slab thicknesses resulted in substantial increase of the $Im(\varepsilon )$ due to the surface state contributions. Independently, these predictions have been confirmed experimentally by Drachev \etal \cite{drachev08} by simultaneous measurements of reflectance and transmittance of differently shaped silver nanostructures. They show that the dramatic increase of $Im[\varepsilon(\omega)]$ for both TE and TM light polarizations is due to the decrease of characteristic dimensions of the silver-based nanostructures; agreeing well with our predictions \cite{zhu08}.

In the present work, we study physical and chemical processes affecting optical losses of silver surfaces that substantially contribute to the optics of metamaterials (the artificially fabricated nano-materials \cite{drachev08}). The adsorption of water, ethanol, and methanol on silver nano-films is studied by first principle density functional theory (DFT). These systems are widely used in modern nanotechnology and nano-physics \cite{morgenstern02,michaelides06,drachev08}. Contribution of the surface states to optical losses is studied by calculations of the imaginary part of the dielectric function for metallic slabs of different thickness. We obtained substantial modifications of the optical function of metallic nano-slabs in the near infrared and visible spectral regions that are discussed in comparison with literature.

\section{Method} 

Typical metallic nanoparticles are much bigger in size than the molecules studied. Consequently the interactions of small molecules with the nano-particles can be studied using widely used model of the surface physics: molecule on an atomic slab \cite{kolasinski08}. Equilibrium atomic structures of silver surfaces with organic molecules are obtained from the total energy minimization method within DFT using {\it ab initio} (gradient-corrected) functionals \cite{delley90, delley00}. The basis functions are given numerically as values on an atomic-centered spherical-polar mesh, rather than as analytical functions (\ie Gaussian orbitals), as implemented in the DMol$^3$ package \cite{matstud}. We employ the super-cell approach to model the surface; our unit cell is a slab consisting of seven monolayers of $Ag~(111)-\sqrt{3}\times\sqrt{3}$, of which the top three layers were allowed to relax. A Monkhorst Pack mesh of $6\times6\times1$ \textbf{k}-points was used to sample the Brillouin zone. Explicit tests were performed to verify that the results were converged for the parameters given above.

The electronic structure and optical properties of the system were then calculated using {\it ab initio} norm-conserving (NC) pseudopotentials (PP), as implemented in the CASTEP package \cite{matstud}. The eigen value problem was solved using the first principle PP method with NC-PP \cite{fuchs99}. Convergence of the results has been carefully checked by choosing energy cut-off values up to 1000eV. The cut-off energy of 550eV was chosen for NC-PP to ensure reliable convergence of predicted optical data. Choice of pseudopotentials (ultrasoft or norm-conserving) does not significantly affect the results. Exchange and correlation (XC) interaction is modeled using the Perdew-Burke-Ernzerhof formalism\cite{perdew96}. A Monkhorst Pack grid of $6\times6\times1$ was used to sample the Brillouin zone, just like for DMol$^3$, yielding well converged results.

Calculated self-consistent eigen energies and eigen functions are used as inputs for optical calculations. Within its penetration depth beneath the surface, the incident light field ($E(\omega)$), at frequency $\omega$, induces a linear optical susceptibility function $\chi^{(1)}$:

\begin{equation}
P_i(\omega)   =  \chi^{(1)}_{ij} E_j(\omega)
\label{equation:1st-order}
\end{equation}

The function $\chi^{(1)}$ is then used to determine the complex dielectric function ($\varepsilon  = \varepsilon_1+ i\varepsilon_2$) through the linear optical susceptibility: $\varepsilon = 1 + 4\pi \chi^{(1)}$. Within the Random Phase Approximation (RPA) the function $\varepsilon$ is given by \cite{gavrilenkoSPIE}:

\begin{equation}
\varepsilon (\omega )=1 - \frac{8\pi }{\Omega _0}\frac{1}{N}\sum _{\bm k}\sum _{c,v}
\frac{{\bm p}_{cv}({\bm k}){\bm p}_{cv}^*({\bm k})}{[E_c({\bm k})-E_v({\bm k})]^2}F_{cv}(\omega ,{\bm k} )
\label{eq:eps}
\end{equation}

\noindent where indices $c$, $v$ run over conduction (empty) and valence (filled) electron states, respectively. The spectral function $F_{cv}(\omega ,{\bm k})$ at the incident light frequency $\omega$ is given by \cite{gavrilenkoSPIE}: 

\begin{eqnarray}
F_{cv}(\omega ,{\bm k}) & = &   \frac{1}{\omega + i\eta -E_c({\bm k}) + E_v({\bm k}) }  \nonumber \\
  & + &  \frac{1}{-\omega - i\eta -E_c({\bm k}) + E_v({\bm k}) } 
\label{eq:Fcv}
\end{eqnarray}

The normalization quantity $\Omega_0 $ has dimension of volume in direct space. The $\varepsilon$ represents the {\it local} part of the dielectric function. In this study we did not include non-localities due to the local field effect.  Note that Eqs. (\ref{eq:eps}) and (\ref{eq:Fcv}) represent an effective value for the dielectric function averaged over different tensor component of $\varepsilon_{\alpha,\beta}$ (where $\alpha, \beta = x, y, z$) in direct space. A more detailed explanation can be found in \cite{gavrilenkoSPIE}.

The stability of various equilibrium configurations is judged by the adsorption energy. The adsorption energy per molecule is defined as the difference between the total energy of the adsorption system and the energies of the isolated components, namely the clean substrate and the adsorbate, divided by the number of adsorbed molecules \emph{N} per unit cell:

\begin{eqnarray}
E_{ads}=\frac{E_{tot}-(E_{substrate}+N{\cdot}E_{adsorbate})}{N}
\label{eq:ads_enrg}
\end{eqnarray}

\section{Results and discussion} \label{sec:results}
\noindent Realistic modeling of optical functions of solid surfaces and interfaces still remains challenging for the first principle theories \cite{downer01}. As a first step it requires reliable reproduction of the ground state of the material (equilibrium atomic geometry) by minimizing the free energy. As a next step optical properties could be calculated using the optimized geometry as an input. Description of excited states, which could be in good agreement with experiment, normally requires inclusion of local field, many-body (excitons) effects, and probably other nonlocal contributions \cite{gavrilenko97,leitsmann05}. This makes theory much more complicated than frequently used independent particles approach (or Random Phase Approximation, RPA). In this work we focus on relative changes of electron energy structure and optical response caused by molecular adsorption on Ag~(111) surface, which could be realistically described without inclusion of many-body effects in optics \cite{gavrilenko06,gavrilenko97}. As such we chose to calculate the dielectric function spectra within the RPA employing norm-conserving pseudopotentials \cite{hamann79}.

\subsection{Equilibrium geometries}

\begin{figure*}[ht]
\centering
\includegraphics[width=0.75\textwidth]{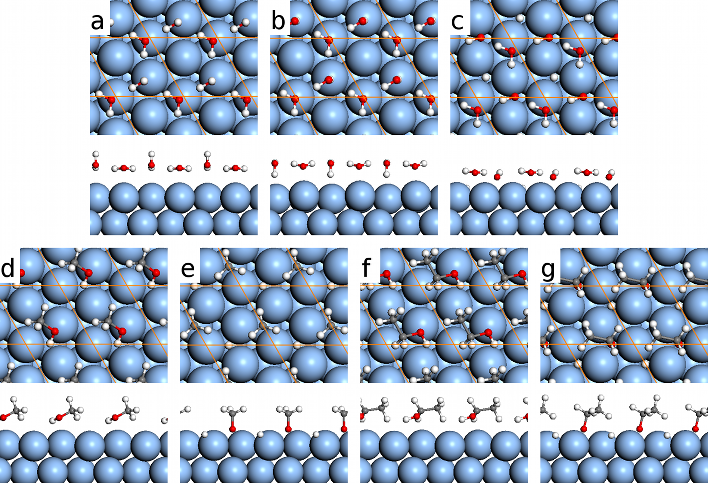} 
\caption{Top and side views of the optimized unit cell configurations of water (a, b, and c), methanol (d and e), and ethanol (f and g) molecules adsorbed on Ag~(111) surface. Panels represent the following configurations: (a)~\ce{H2O}-up,  (b)~\ce{H2O}-down, (c)~\ce{H2O}-down-dsc, (d)~\ce{CH3OH}, (e)~\ce{CH3OH}-dsc, (f)~\ce{CH3CH3OH}, and (g)~\ce{CH3CH2OH}-dsc. The dissociate (dsc) configurations all have one hydrogen atom desorbed from the respective molecule and enter into a reaction with the surface.}
\label{fig:Geom-Mol}
\end{figure*}

\begin{table*}
\caption{Chemisorption and physisorption adsorption energies (as calculated by Eqn. \ref{eq:ads_enrg}) and optimized structural parameters of \ce{H2O}, \ce{CH3OH}, and \ce{CH3CH2OH} adsorbed on the Ag~(111) surface. The average surface height is used for the \ce{O\sbond Ag} and \ce{H\sbond Ag} distances, representing the distance between the surface and the oxygen/hydrogen. Since there are two \ce{H2O} molecules per unit cell, two distances are provided. $\Delta$Ag is the difference in the average height of the top silver layer compared to a bare surface. \ce{O\sbond H}, \ce{O\sbond O}, and \ce{O\sbond C} represent the interatomic distances in the system; multiple distances are provided as needed.}
\begin{ruledtabular}
\begin{tabular}{l c c c c c c c}
Structure & $E_{ads}$ (eV) & \ce{O\sbond Ag} (\AA) & \ce{H\sbond Ag} (\AA) & $\Delta$Ag (\AA) & \ce{O\sbond H} (\AA) & \ce{O\sbond O} (\AA) & \ce{O\sbond C} (\AA)\\
\hline
\ce{H2O}-up			& $-0.9047$ & $2.69, 2.95$ 	& --- 	& $+0.0007$	& $0.978, 0.996, 1.01$	& $2.78, 2.95, 2.98$	& ---\\
\ce{H2O}-down		& $-0.9471$ & $2.64, 3.18$ 	&$2.06$	& $+0.0155$	& $0.994, 0.998, 1.01$ 	& $2.84, 2.92, 2.94$	& ---\\
\ce{H2O}-down-dsc	& $-0.1876$ & $1.83, 2.25$	&$0.846$	& $+0.0233$	& $0.986, 0.991, 1.06$	& $2.51, 2.94, 3.37$	& ---\\
\ce{CH3OH}			& $-0.6017$ & $2.51$			&$2.24$	& $+0.0017$	& $0.985$				& $5.00$				& $1.42$\\
\ce{CH3OH}-dsc		& $+0.0622$ & $1.46$			&$0.968$	& $+0.0375$	& $2.94$					& $5.00$				& $1.39$\\
\ce{CH3CH2OH}		& $-0.6694$ & $2.61$			&$2.34$	& $-0.0026$	& $0.984$				& $5.00$				& $1.42$\\
\ce{CH3CH2OH}-dsc	& $+0.0004$ & $1.48$			&$0.911$	& $+0.0272$	& $2.96$					& $5.00$				& $1.40$\\
\end{tabular}
\end{ruledtabular}
\label{tbl:Geom}
\end{table*}%

Table~\ref{tbl:Geom} lists the adsorption energies $E_{ads}$, per molecule, as calculated by Eqn.~\ref{eq:ads_enrg} and the optimized structural parameters of the adsorbates on the surface. Atomic relaxation of water on transition metals has been studied both theoretically and experimentally \cite{methfessel92,michaelides03, michaelides06, meng04, morgenstern02}. Two different atomic geometries of water molecules physisorbed on Ag~(111) surface are predicted: with hydrogen up and down. Our predicted geometries of physisorbed water molecules on Ag~(111) surface agree, for the most part, well with those reported in Refs.~\onlinecite{michaelides03,michaelides06} where a similar theoretical framework was used. The \ce{H2O}-up and \ce{H2O}-down geometries are shown in Fig.~\ref{fig:Geom-Mol}a and b, respectively. 

We also studied the effect of water molecule dissociation on optical properties of silver. At room temperature the reaction \ce{H2O <=> HO- + H+} occurs spontaneously. The proton is then free to enter into reaction with the surface. For a single water molecule adsorbed on a closed-packed Ag~(111) surface the distance between the oxygen and the host Ag atoms, the \ce{O\sbond Ag} distance, of 2.78 {\AA} was reported in Ref.~\onlinecite{michaelides03}. Once enough \ce{H2O} molecules are added to the system to form hexamers on the surface, the molecules form an ice-like structure. For a single bilayer of \ce{H2O} adsorbed on the surface, the zig-zag configuration of ice isn't as pronounced, as for water molecules not in contact with the surface. Our oxygen-silver surface distances can be found in Table~\ref{tbl:Geom}. The water hexagons are slightly irregular as can be seen in Figs.~\ref{fig:Geom-Mol}a,b. The shift is much more pronounced in the \ce{H2O}-up configuration. This asymmetry of the hexagonal structure of water molecules was previously reported for Ru\{0001\} \cite{michaelides03}. Once one of the hydrogen is desorbed, see Fig.~\ref{fig:Geom-Mol}c, the remaining \ce{OH} molecule shifts much closer to the surface and the water hexagonal structure becomes even more irregular. 

Though the average surface height doesn't change very much, see $\Delta$Ag in Table~\ref{tbl:Geom}, the individual silver atoms on the surface do move quite a bit more. This can be seen clearly in the side view of Fig.~\ref{fig:Geom-Mol}b. Another observation is that the water molecules that do not lose their hydrogens are not oriented parallel to the surface, they are tilted with an angle of 7.69\degree. We report a negative adsorption energy of -0.1876eV, as predicted by DMol$^3$, indicating the stable structure. This is contrary to what was reported in Refs.~\onlinecite{meng04,michaelides03}. Previously, we carefully compared equilibrium geometries predicted by CASTEP and DMol$^3$ with different molecules\cite{gavrilenko06,agavrilenko08} and showed that the use of localized basis in DMol$^3$ is beneficial in terms of prediction of stable geometries. Consequently equilibrium molecule-surface distances predicted by LDA DMol$^3$ method better agree with those generated by VASP package including van der Waals contributions \cite{ortmann05} and agree much better with experiment.

Methanol (\ce{CH3OH}, Fig.~\ref{fig:Geom-Mol}d) and ethanol (\ce{CH3CH3OH}, Fig.~\ref{fig:Geom-Mol}f) adsorb very similarly on the surface of silver (see Table~\ref{tbl:Geom}). Once the O-H bond is broken, the molecules rotate to bring the oxygen closer to the surface, however, the molecules also shift on the surface and position themselves such that the oxygen is between three Ag surface atoms. It is important to note that although a stable configuration was found with the O-H bond broken for both methanol and ethanol, those configurations are metastable at best. Both methanol and ethanol (\ce{CH3OH}-dsc and \ce{CH3CH2OH}-dsc) have positive adsorption energies. One can speculate however, that the van der Waals force correction may remarkably modify the adsorption energies \cite{ortmann05} thus yielding stable configurations for dissociated configurations of both methanol and ethanol.

\subsection{Dielectric function and optical losses}
Optical losses of a system are proportional to the imaginary part of dielectric susceptibility function, ($Im(\varepsilon)$)\cite{saleh07,sipe96,kolasinski08}. Here we discuss the $Im(\varepsilon)$ spectra calculated according to Eqs.~\ref{eq:eps} and \ref{eq:Fcv} for 7 monolayer thick Ag~(111) slab with and without different adsorbed molecules.  We also compare these results with the experimental data for bulk Ag\cite{j-cr72}. 

\begin{figure}[ht]
\centering
\includegraphics[width=0.95\columnwidth]{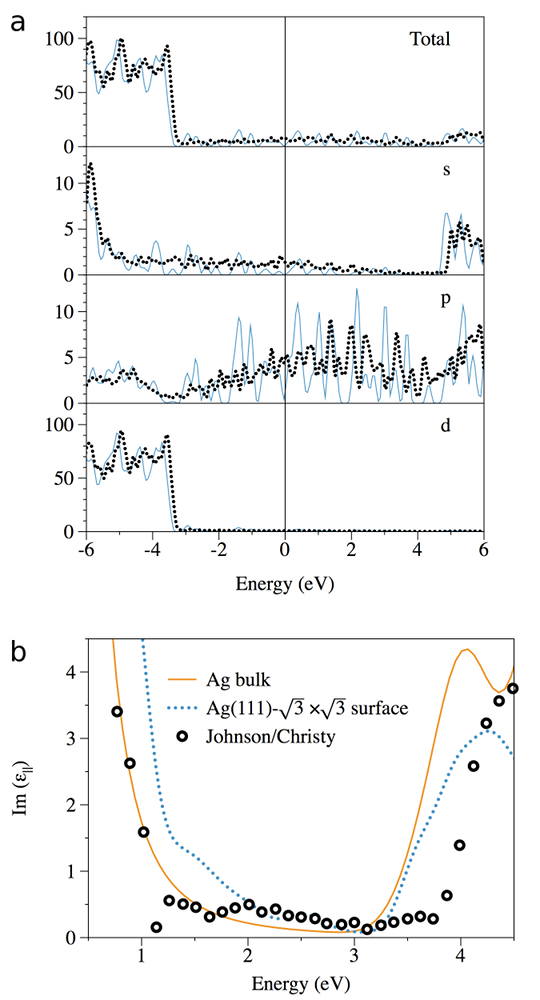}
\caption{(a)~Projected density of states of bulk (solid blue) and 7 monolayer thick slab (dotted black) of silver. The shift in the \emph{d}-orbital is due to the redistribution of electrons. (b)~Calculated spectra of immaginary part of dielectric susceptibility function for bulk (solid orange) and 7 monolayer thick Ag~(111) slab (dotted blue) in comparison with experimental data measured in Ref.~\onlinecite{j-cr72} (circles).}
\label{fig:PDOS_DF_Ag}
\end{figure}

First, the predicted optical spectra for bare Ag~(111) surface are considered.  Without quasi-particle correction the predicted $Im(\varepsilon)$ LDA data calculated for a slab with experimental lattice constant shows a remarkable red shift in comparison with experiment. The LDA method used in this work produces compressed lattice of bulk Ag. Effect of lattice compression on electron energy structure results in blue shift of optical spectra \cite{martin04} thus improving a comparison with experiment as demonstrated in Fig.~\ref{fig:PDOS_DF_Ag}.  

Optical excitations from the surface states are responsible for the calculated increase of  $Im(\varepsilon)$ of Ag-slab in visible and near infrared optical range with respect to bulk (see $Im(\varepsilon)$ data between 1 to 2.5 eV in Fig.~\ref{fig:PDOS_DF_Ag}).

\begin{figure*}
\centering
\includegraphics[width=\textwidth]{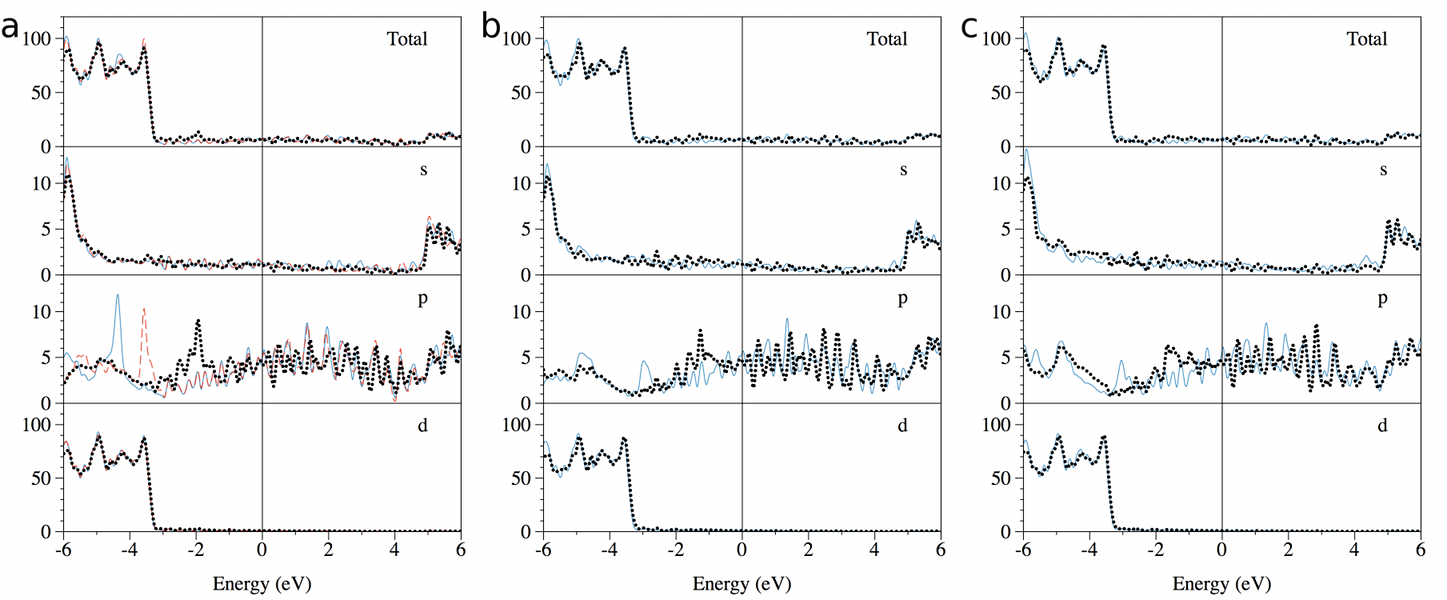}
\caption{Projected density of states spectra of \ce{H2O} (a), \ce{CH3OH} (b), and \ce{CH3CH2OH}(c)  adsorbed on Ag (111) film. The thin (blue and red) lines depict the physisorbed configurations of (a)\ce{H2O}-up (blue) and \ce{H2O}-down (red), (b)\ce{CH3OH}, and (c)\ce{CH3CH2OH}; the dotted black line represents the chemisorbed configurations. The upper panel shows the total DOS and the other panel present orbital projected DOS as indicated by the panel indexes. The PDOS of bulk and silver surface can be found in Fig.~\ref{fig:PDOS_DF_Ag} for comparison.}
\label{fig:PDOS}
\end{figure*}

In order to understand the nature of predicted changes in the spectra of $Im(\varepsilon)$ the projected density of states (PDOS) spectra are calculated for the systems studied. Details of the calculations of PDOS are described in Ref.~\onlinecite{gavrilenkoSPIE}. Our previous analysis of calculated band structure and PDOS of the Ag-slab (without molecules) \cite{zhu08} indicated substantial contributions of surface electron energy states which are in resonance with the bulk electron states and caused redistribution of the resulting density of states.  In this paper we present the PDOS spectra analysis and focus on the effects on PDOS caused by the molecular adsorption. 

Optical transitions between the occupied states and Fermi surface are responsible for the predicted substantial increase of the $Im(\varepsilon)$ spectra of Ag~(111) surface as compared with bulk in the near infrared region as demonstrated in Fig.~\ref{fig:PDOS_DF_Ag}. The transitions between the occupied states near $-1.6$ eV and Fermi level are responsible for the calculated shoulder at 1.55 eV in $Im(\varepsilon)$ spectra of bare surface, as shown in Fig.~\ref{fig:PDOS_DF_Ag}. 

\begin{figure}[ht]
\centering
\includegraphics[width=0.95\columnwidth]{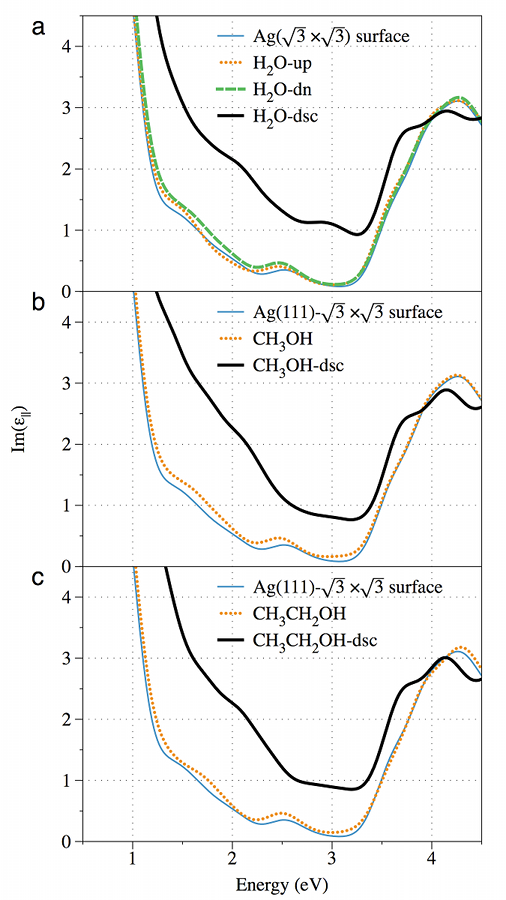} 
\caption{Imaginary part of calculated dielectric function of water (a), methanol (b), and ethanol (c) adsorbed on Ag~(111)-$\sqrt{3}\times\sqrt{3}$ surface. The thin, dotted and dashed lines represent the undissociated configuration, while the bold solid lines correspond to dissociated molecules on the surface, respectively.}
\label{fig:DF_All}
\end{figure}

Crystal truncation results in a predicted redistribution of charge between 4\emph{d}- and 5\emph{s}-orbitals causing effective energy shift of the 4\emph{d}-orbitals towards lower energies (see Fig.~\ref{fig:PDOS_DF_Ag}). Based on the PDOS data analysis we can state that predicted modifications of the $Im(\varepsilon)$ spectra of Ag~(111) surface with respect to bulk are caused by surface-to-bulk and bulk-to-surface optical transitions. This conclusion is in the alley of previously reported analysis \cite{monachesi01}, thus confirming bulk-to-surface nature of relevant electron transitions. Optical features predicted on Ag~(111) above 3.5 eV are characterized by bulk electron transitions with minor contributions of the surfaces electrons.

Molecular adsorption substantially modifies spectra of $Im(\varepsilon)$. In this work we study the effect of molecular dissociation on the Ag~(111) surface. At room temperature most solutions contain dissociated molecules which could interact with the surface states \cite{kolasinski08}. In Fig.~\ref{fig:PDOS} we present the effects of molecular adsorption on the calculated PDOS spectra of a Ag~(111) film. On the bare Ag~(111) surface the electron energy structure is substantially modified as compared to the bulk Ag. Molecular adsorption on the crystal surface and atomic structure relaxation resulted in substantial overal increase of DOS related to the 5\emph{s}- and 4\emph{p}-orbitals as shows in Fig.~\ref{fig:PDOS}. We observed at least two pronounced PDOS 5\emph{s}-peaks located near $-0.4$ and $-1.6$ eV (occupied states) on Ag~(111) surface as well as energy increase of both 5\emph{s}- and 4\emph{p}-unoccupied states in the region between 0 and 4 eV.  No substantial differences between surface and bulk of the 4\emph{d}-unoccupied PDOS has been observed. 

Analysis of the PDOS spectra shows that changes of molecular orientations on Ag~(111) surface result in modifications of mostly unoccupied \emph{p}-electron states around 1.6 eV and higher. The $Im(\varepsilon)$ function spectra calculated in visible and near infrared spectral regions are shown in Fig~\ref{fig:DF_All}.

 In order to understand the effect of molecular dissociation on optical losses of the Ag~(111) surface we calculated molecular configurations after the dissociation of a hydrogen atom from all molecules considered here. Geometry optimization study indicates that the dissociated H-atom tends to interact much stronger with the surface. The same effect for water adsorbed on Ag~(111), was reported earlier in Refs.~\onlinecite{michaelides03,michaelides06}. According to our total energy minimization study, the dissociated hydrogen atom moved to the interstitial region. The adsorbed water pattern remarkably shifts while retaining the ice-like hexagonal structure. Dissociation caused enhancement of the interactions between unpaired oxygen and hydrogen electrons. In all cases the dissociation of hydrogen leads to further substantial increase of $Im(\varepsilon)$ function which means the increase of optical losses of Ag~(111) in relevant spectral regions. 

Comparative analysis of the calculated PDOS spectra of Ag~(111) surface with dissociated molecules and that of bulk Ag show strong modifications of the electron density: appearance of new occupied states located at around $-1.7$ to $-2.5$ eV below Fermi energy and unoccupied states near 0.5 to 0.7 eV above the Fermi level. Consequently in dissociative configuration there are contributions to the $Im(\varepsilon)$ function of electron transition - between occupied surface states and Fermi surface and between occupied and unoccupied surface states. The corresponding energy separations are between 1.5 to 3.2 eV. Contributions of the transitions between Fermi surface and unoccupied host states to the observed changes seem to be minor. Our results therefore indicate that chemical reactions between dissociated H- and O-atoms occur mostly with unpaired 5\emph{s}- and 4\emph{p}-surface electrons of the Ag-atoms. 

Predicted strong increase of optical losses in infrared and visible range apparently dominates over that calculated in the near UV region (see Fig.~\ref{fig:DF_All}). According to the PDOS spectra the \emph{d}-electrons are involved in optical excitations in the region above 3.2 eV. In contrast to the infrared and visible region, the molecular adsorption in ultraviolet somewhat even reduces optical losses in the region above 4.0 eV (see Fig.~\ref{fig:DF_All}) due to the discussed charge redistribution in the near-the-surface region. These findings should be interesting for the optical engineering of the UV devices.   

Analysis of the data presented in Figs.~\ref{fig:PDOS_DF_Ag} and \ref{fig:DF_All} indicates several features related to the effect of molecular adsorption on the Ag~(111) surface. Physisorption only slightly modifies the $Im(\varepsilon)$ spectra mostly due to minor changes of atomic geometries. The most substantial observed modifications of the $Im(\varepsilon)$ spectra are caused by chemisorption: molecular dissociation and creation of new inter-atomic bonds. Adsorption of molecules substantially enhances optical losses in infrared region. Comparison with PDOS spectra indicate redistribution of the charge associated with 5\emph{s} and 4\emph{p} silver orbitals (see Fig.~\ref{fig:PDOS}).

Results of this study clearly demonstrate strong contribution of the surface states to the optical losses on Ag~(111) surface.  Adsorption of organic molecules (ethanol and methanol) result in further increase of the losses on Ag~(111) surface. Nature of the observed increase of the $Im(\varepsilon)$ function of  Ag~(111) surface induced by molecular adsorption is substantial modifications of surface electron energy structure and dramatic enhancement of electron transitions between surface and bulk states.

\section{Conclusions}
Equilibrium atomic structure, projected density of states, and imaginary part of optical dielectric function spectra of Ag~(111) surface covered with water, methanol, and ethanol molecules were studied using first principle density functional theory. Strong contribution of the surface states on optical losses is observed. Substantial modifications of optical functions of metallic nano-slabs in near infrared and visible spectral regions, caused by surface states and molecular adsorption is predicted. Our results indicate increase of the imaginary part of dielectric function of the Ag~(111) surface due to charge redistribution of  the surface Ag-atoms accompanied  by the enhancement of electron transitions probabilities between surface and bulk states. 

\section{Acknowledgments} 
Authors acknowledge helpful discussions with M. A. Noginov, V. M. Shalaev, and V. P. Drachev. This work is supported by NSF PREM DRM-0611430 and NSF NCN EEC-0228390

\bibliographystyle{apsrev4-1}

%

\end{document}